\documentclass{sigchi-ext}
\usepackage[T1]{fontenc}
\usepackage{textcomp}
\usepackage[scaled=.92]{helvet} 
\usepackage{graphicx} 
\usepackage{balance}  
\usepackage{booktabs} 
\usepackage{ccicons}  
\usepackage{ragged2e} 

\def\plaintitle{SIGCHI Extended Abstracts Sample File: Note Initial
  Caps} 
\def\emptyauthor{}
\def\plainkeywords{Smart speaker; tangible privacy; privacy hat; iot study platform}

\title{Making Privacy Graspable: Can we Nudge Users to use Privacy Enhancing Techniques?}

\numberofauthors{4}
\author{%
  \alignauthor{%
    \textbf{Christian Tiefenau}\\
    \affaddr{University of Bonn} \\
    \affaddr{Bonn, Germany} \\
    \email{tiefenau@cs.uni-bonn.de} 
  }
  \alignauthor{%
    \textbf{Maximilian H{\"a}ring}\\
    \affaddr{Fraunhofer FKIE}\\
    \affaddr{Bonn, Germany}\\
    \email{haering@cs.uni-bonn.de} 
  } \vfil 
  \alignauthor{%
    \textbf{Eva Gerlitz}\\
    \affaddr{Fraunhofer FKIE}\\
    \affaddr{Bonn, Germany}\\
    \email{gerlitz@cs.uni-bonn.de} 
  } 
  \alignauthor{%
    \textbf{Emanuel von Zezschwitz}\\
    \affaddr{University of Bonn, Fraunhofer FKIE}\\
    \affaddr{Bonn, Germany}\\
    \email{zezschwitz@cs.uni-bonn.de} 
  }
}

\definecolor{linkColor}{RGB}{6,125,233}
\hypersetup{%
  pdftitle={\plaintitle},
  pdfauthor={\emptyauthor},
  pdfkeywords={\plainkeywords},
  bookmarksnumbered,
  pdfstartview={FitH},
  colorlinks,
  citecolor=black,
  filecolor=black,
  linkcolor=black,
  urlcolor=linkColor,
  breaklinks=true,
}

\begin{document}

\copyrightinfo{Copyright is held by the author/owner. Permission to make digital or hard copies of all or part of this work for personal or classroom use is granted without fee. Poster presented at the 15th Symposium on Usable Privacy and Security (SOUPS 2019).}

\begin{marginfigure}[15pc]
    \includegraphics[width=\marginparwidth]{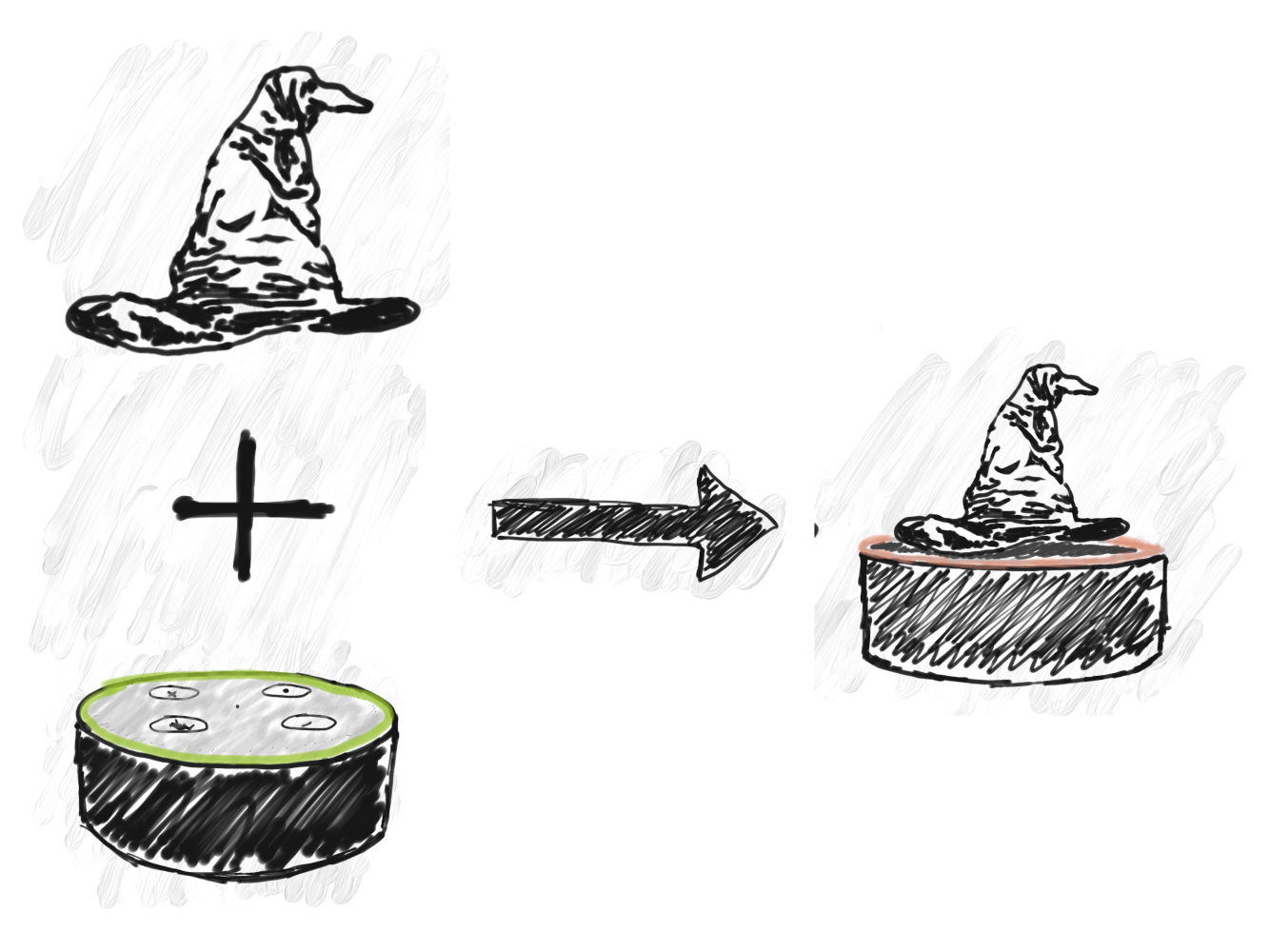}
    \caption{Concept of the \textit{Privacy Hat}: Muting a smart speaker by placing an object on top of it.}
    \label{fig:prototypeidea}
\end{marginfigure}
\maketitle

\RaggedRight{} 

\begin{abstract}
Smart speakers are gaining popularity. However, such devices can put the user's privacy at risk whenever hot-words are misinterpreted and voice data is recorded without the user's consent. To mitigate such risks, smart speakers provide privacy control mechanisms like the build-in mute button. Unfortunately, previous work indicated that such mute buttons are rarely used. In this paper, we present the \textit{Privacy Hat}, a tangible device which can be placed on the smart speaker to prevent the device from listening.
We designed the \textit{Privacy Hat} based on the results of a focus group and developed a working prototype. We hypothesize that the specific user experience of this physical and tangible token makes the use of privacy-enhancing technology more graspable for the user. As a consequence, we expect that the \textit{Privacy Hat} nudges users to more actively use privacy-enhancing features like the mute button. In addition, we propose the \textit{Privacy Hat} as a study tool as we hypothesize that the artifact supports participants in reflecting their behaviour. We report on the concept, the prototype and our preliminary results.
\end{abstract}



\section{Introduction}
The number of smart speakers like the Amazon Echo \cite{AmazonEcho}, the Google Home \cite{GoogleHome} or Apple's HomePod \cite{AppleHomePod} grew 2018 by 39.8\% in the US, resulting in an estimated usage by 66.4 million American households \cite{Voicebot}.
Since their introduction in 2014, usable security research aimed at understanding usage behavior and risk perception of smart speaker users. Several studies found that users generally express little privacy concerns, especially due to an incomplete understanding of the security risks or resignation \cite{bugeja2016privacy, Ammari:2019:MSI:3328720.3311956, lau2018alexa}. Privacy controls like the mute button or audio logs seem to be rarely used. One reason for these circumstances can be that these mechanisms do not align with the users' privacy control needs. Related work suggests to nudge users with privacy notices to reflect on decisions, so that they can take informed privacy-related actions \cite{lau2018alexa}. Following this line of thoughts, we hypothesize that providing users with a physical and tangible device can help them to a better understanding of the privacy-topic and furthermore nudge them to reflect on their daily smart speaker interaction.

\begin{marginfigure}[-20pc]
    \includegraphics[width=\marginparwidth]{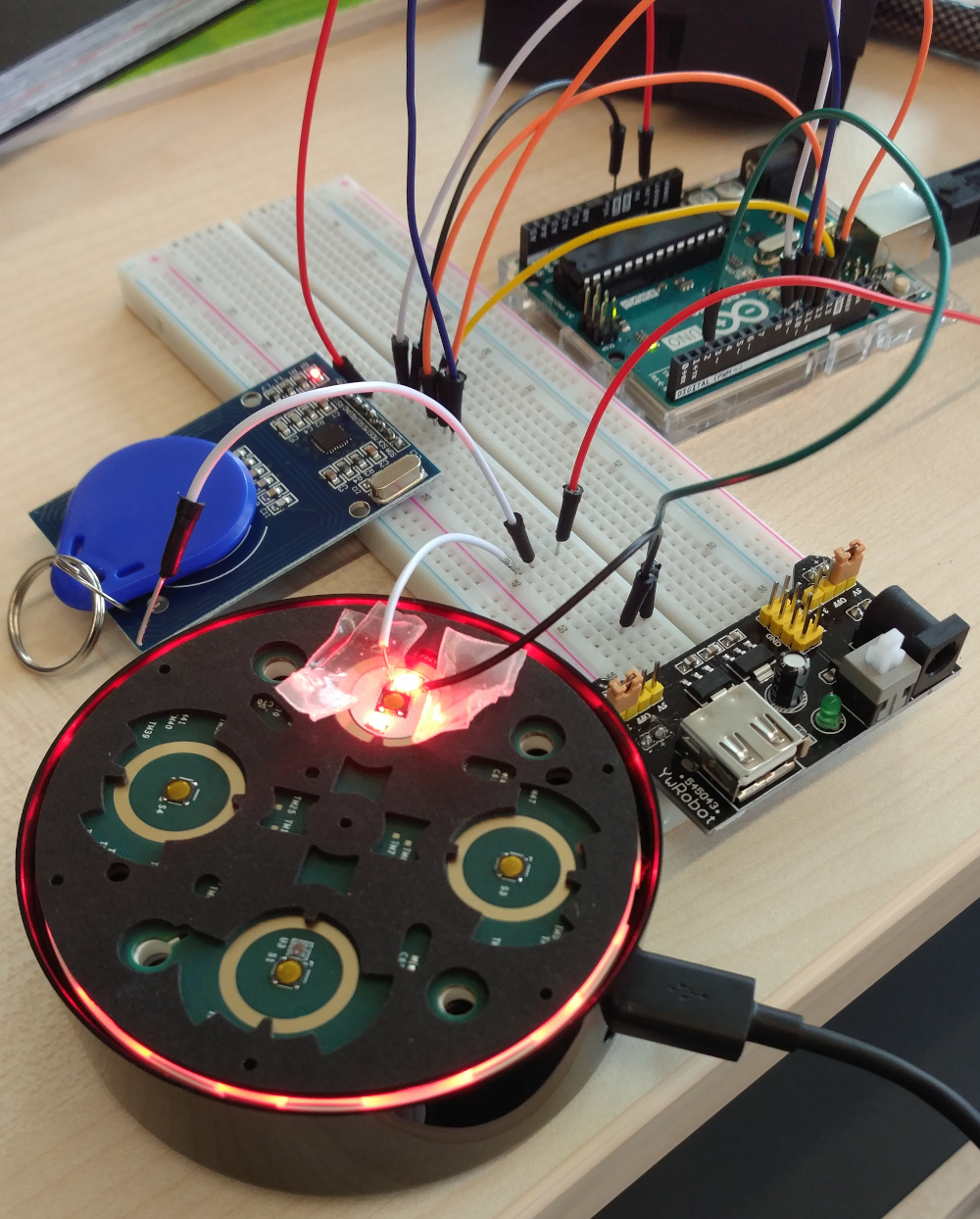}
    \caption{Our prototype in the early stage. Here, we tried using a RFID-tag scanner in combination with an Arduino.}
    \label{fig:prototyping}
\end{marginfigure}

Closely related to our project is the artistic DIY-concept ``Project Alias'' \cite{ProjectAlias}. Users can 3D-print a physical token that is then placed on top of a smart speaker. The use of white noise disables the speakers' ability to listen to the activation word. The device itself serves as a trusted component which can be activated by an individual hot word. If the hot word is received, ``Project Alias'' stops sending white noise and activates the smart speaker for further interaction. While we acknowledge ``Project Alias'' as a potential solution, it does not foster physical interaction with the device itself nor does it communicate the device's status. 

In contrast, our \textit{Privacy Hat} is a tangible object that can be placed on top of a smart speaker to prevent it from listening and was designed to provide implicit visual feedback on the smart speaker's state. Since anyone in the room can observe if the \textit{Privacy Hat} is placed on the device or not, we expect that the concept can provide unintrusive but clear privacy notices that may trigger usage-reflection. To inform the design of \textit{Privacy Hat}, we performed a focus group with four smart speaker users. We asked them about their current smart speaker usage behavior and discussed alternative interaction paradigms that could support privacy-awareness and trigger the active use of the privacy control ``muting.''

In addition to the development of the \textit{Privacy Hat}, we present a smart home study-framework that enables us to get feedback of the participants' actual device usage in the field. This framework will be used to measure the effect of the \textit{Privacy Hat} in our participants' actual smart homes.

We like to note that our concept is not simply a usability-optimized solution to the privacy-related problems of smart-home users. For this we would have proposed a concept which supports remote control of such a functionality. On the contrary, our concept suggests that users have to actively put an object on top of the device and thus have to physically go to the speaker to perform the action. We hypothesize that requiring explicit physical interaction makes the process more tangible, has potential to affect the users' privacy-related behavior and helps them to reflect about their behavior. We anticipate that this tool will give us additional insights into the smart home user's reasoning when using privacy-enhancing technology.
\newpage

\subsection{Research questions}
With the concept and preliminary study presented in this paper, we aim to explore the following research questions: 
\begin{itemize}
    \item How do users interact with the traditional mute button and what do they think about it?
    \item Is a tangible ``Mute-Token'' perceived beneficial by users?
    \item Can a tangible ``Mute-Token'' help gather additional insights on user behavior and thoughts?
\end{itemize}



\section{Privacy Hat}
By providing a new mechanism, so that the act of muting the smart device is transformed into a explicit, physical action, we aim to nudge the user to reflect on their interaction with the device, especially their muting behaviour. In our study this mechanism takes the form of the \textit{Privacy Hat}: Covering the device with the \textit{Privacy Hat} (see Figure \ref{fig:prototypeidea}) will mute it and give a new visual representation on the status of the speaker, different to the usual flashing red LEDs.

\begin{marginfigure}[-20pc]
    \includegraphics[width=\marginparwidth]{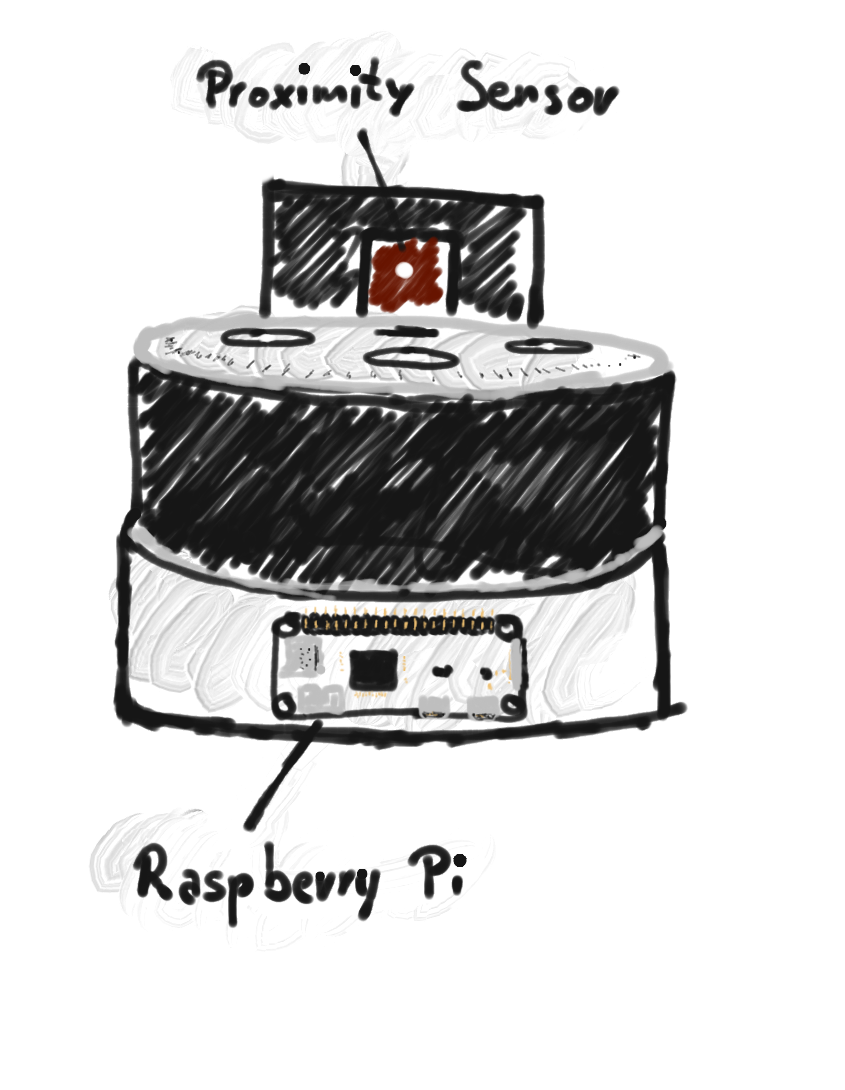}
    \caption{A sketch, how our study-framework looks like. The Raspberry Pi is placed below the Echo Dot and connected internally to its mute button and LED. The distance sensor is placed on the side of the platform to detect objects on top of the device.}
    \label{fig:prototype}
\end{marginfigure}

To test this concept and before we developed the prototype, we were interested in users' opinions towards the idea of using this way to mute a smart speaker and if it would change their perception towards the device. 
For this, we conducted a focus group with four participants who had already used an Amazon Echo in their households for several tasks, including setting a timer, listening to music or news and controlling smart home devices like smart bulbs or outlets. They positively stressed its sound and ``human''-like voice while disliking when they had to talk about the device itself whenever other people are visiting them in their homes.
All of them stated that privacy is important to them. One participant regularly used the mute button before going to bed and one mentioned using it when talking about sensitive data.

After introducing them to the principle of the \textit{Privacy Hat}, we asked them how they would use such a token and how it would change their perception.
They mentioned, that the positioning of the smart speaker is crucial to this concept, so that anyone who is entering the room can see it. In addition to just placing something on top of the device, they presented the idea of ``remote'' muting stations that are placed at several places in the room. 
Also they could imagine using the token as some kind of authentication token, placing it the other way around on top of the speaker for specific actions that require user authentication, for example placing an order. Related to the design they agreed on using some kind of hat that covers the device completely to connote a more secure feeling.

\section{Proposed study design}
For an evaluation of the concept, a platform that enables us to track the muting-behavior of a smart speaker is needed. In the following we will discuss a prototype and give further details on the upcoming study.

\subsection{Prototype}
\begin{marginfigure}[-35pc]
    \includegraphics[width=\marginparwidth]{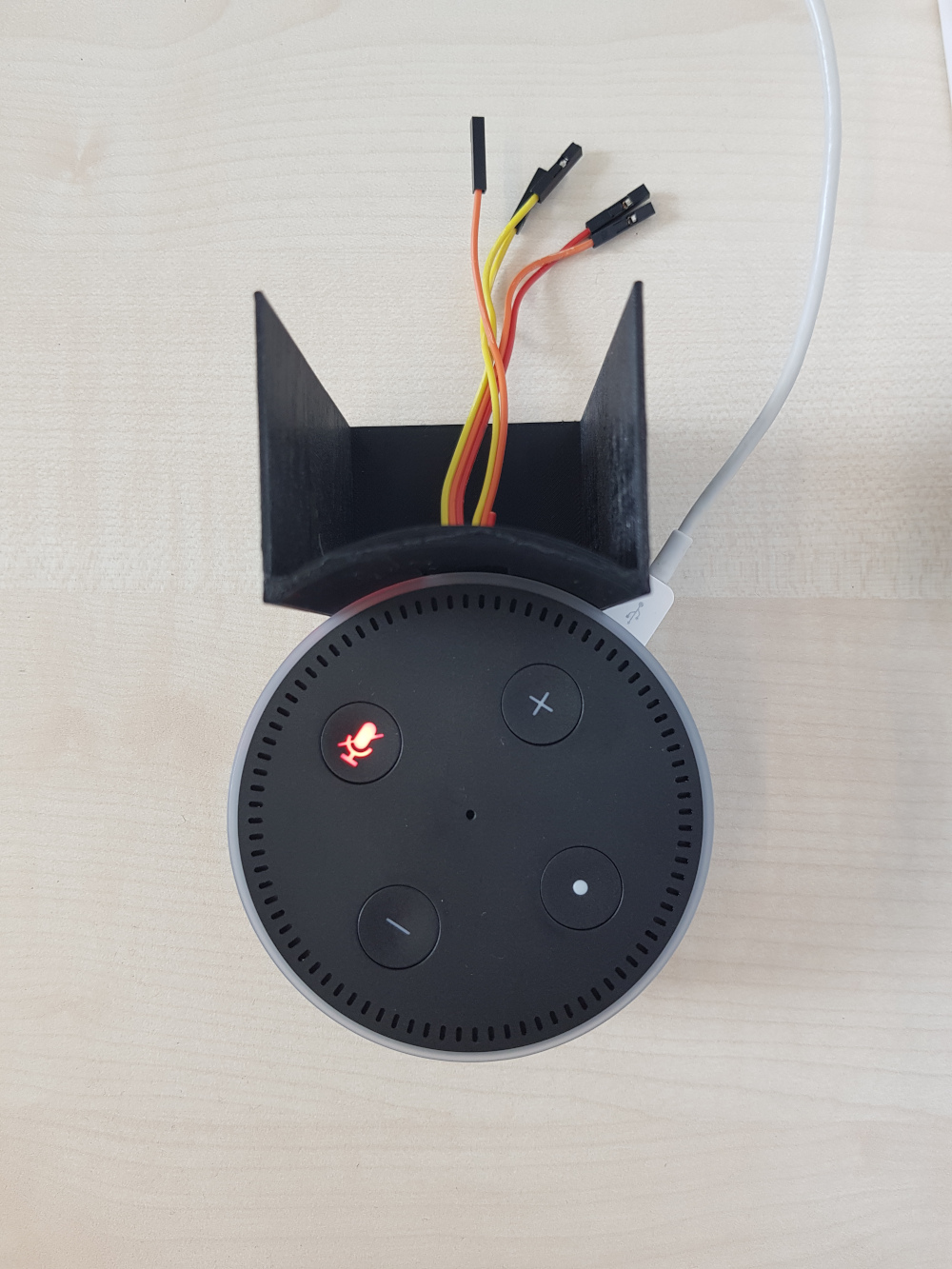}
    \caption{Top view of our prototype with the sensor placed on the side.}
    \label{fig:device}
\end{marginfigure}

We assembled a prototype that uses a Raspberry Pi as the main component. For the smart speaker we chose a modified Amazon Echo Dot as seen in Figure \ref{fig:prototyping}. For the modification, we soldered a wire to the red status LED of the muting button to capture the current muting status of the Echo Dot. Another wire was placed on the contact of the physical mute-button enabling the Raspberry Pi to trigger it. To sense if something is put on the Echo Dot we tried to integrate a distance sensor into the Echo Dot itself. However, this did not turn out as an applicable scenario due to size restrictions. Therefore, we placed the sensor in a 3D-printed case, that acts as a docking station for the Dot (see Figure \ref{fig:device}).
A python script on the Raspberry Pi tracks the muting events of the participants as well as whether the event was triggered by pressing the button or by putting the \textit{Privacy Hat} on the device. The way of muting, the status of the device and the time are then logged.


\subsection{Directions for Future Work}
We are in preparation of a pilot study, that will provide the participants with a default object to mute the Echo Dot, as proposed by the participants of our focus group. However, technically every object could be used to provide this functionality. If this approach turns out to be promising, further studies can evaluate the "best" and most useful design for such a token.
To test if the proposed prototype can help to answer our research questions, we plan to conduct a field study that tests both aspects. Over the course of several weeks, we will gather data and observe muting behavior after introducing the \textit{Privacy Hat}. For this, we will hand out modified Amazon Echo Dots to participants that have not used a smart home speaker before. This way, we try to ensure that the novelty effect of the device will be nearly the same for everyone. 
We want to point out that the novelty effect may benefit our study design. However, we do not want to measure the behavior of users in the most realistic scenario but trigger reflection and experimenting.

The current plan of the study spans 4 weeks and is split into two phases that will each last two weeks to counter the possible novelty effect of introducing a smart home speaker and a resulting higher frequency in usage. 
In phase one we will only be observing the actual usage of the Echo Dot with focus on the muting-behavior. 
For the second phase, which will also last two weeks, we will present the \textit{Privacy Hat} to the participants as an alternative way to mute the device. We will still be collecting the muting behavior of our participants to then evaluate the actual effect of a tangible muting option.

After the two phases we will conduct interviews concerning the participants general opinion about the concept of the \textit{Privacy Hat}, which aspects they like and whether they see potential for improvement. Additionally, we are interested in the participants' perceptions regarding privacy, whether these perceptions change when having a tangible muting option and if they consciously note a shift in their own muting behaviour. For this, we will ask for the muting frequency and which events caused them to use the \textit{Privacy Hat}.

We proposed and discussed the \textit{Privacy Hat}, an example for a graspable privacy enhancing technology in smart homes. We argue that this principle can be used to evaluate privacy perceptions in other scenarios as well. 
We are confident that making the usage of IoT devices more tangible is a promising area worth looking at, since it could help the users to reduce resignation and make them feel more in control.
In addition, the proposed study framework can be used to track actual user behavior in the field and now enables us to compare the already self-reported data with empirical measured ones.

\balance{} 

\bibliographystyle{SIGCHI-Reference-Format}
\bibliography{sample}

\end{document}